\documentclass[twocolumn,showpacs,aps,pre,nofootinbib]{revtex4}

\newcommand{\be}{\begin{equation}}
\newcommand{\ee}{\end{equation}}

\newcommand{\om}{\omega}

\newcommand{\bra}{\langle}
\newcommand{\ket}{\rangle}

\newcommand{\non}{\nonumber}
\newcommand{\bea}{\begin{eqnarray}}
\newcommand{\eea}{\end{eqnarray}}

\begin{document}

\title{Damped harmonic oscillator: pure states of the bath  \\and 
exact master equations 
}
\author{Andrey Pereverzev}
\email{andrey.pereverzev@trinity.edu}
\affiliation{Department of Chemistry, Trinity University, San Antonio, 
TX 78212}

\date{\today}

\begin{abstract}

Time evolution  of a harmonic
oscillator linearly coupled to a heat bath is compared for three classes 
of initial states for the bath modes - grand canonical ensemble, number 
states and coherent states. 
It is shown that for a wide class of number states the behavior of the 
oscillator is similar to the case of the equilibrium bath. 
If the bath modes are initially in  coherent states, 
then the variances of the oscillator coordinate and momentum, as well as 
 its entanglement to the bath,
asymptotically approach the same values as for 
the  oscillator at zero temperature and the average coordinate and 
momentum show a Brownian-like behavior.
We derive an exact master equation for the characteristic function 
of the oscillator  valid for arbitrary factorized initial conditions.
In the case of the equilibrium bath this equation reduces to an equation 
of the Hu-Paz-Zhang type, while for the coherent states bath it leads to  
an exact stochastic master equation with a multiplicative noise.
\end{abstract}

\pacs{05.30.-d, 05.40.-a, 02.50.-r}

\maketitle

\section{Introduction} \label{intro}
The model of an oscillator linearly coupled to the bath of harmonic 
oscillator has played an important part in statistical 
mechanics \cite{Caldeira,Haake,Grabert2}, 
quantum optics \cite{Walls,Carmichael} and  
quantum measurement theory \cite{Unruh,Hu,Guilini}. 
In most studies of this model the initial state for the whole system is 
taken as a mixed density matrix. In particular, one often uses factorized 
initial state where the bath modes are in thermal equilibrium and 
the oscillator is in a pure state. Less often the so-called thermal 
initial conditions are used \cite{Hakim,Grabert,Romero,daCosta}. Pure states 
of the bath are rarely considered \cite{Petrosky}, except for the vacuum state 
of the bath. 

Our goal in this paper is to compare the behavior of the oscillator for 
different pure initial states for the bath modes to the case of the 
bath in equilibrium. The quantities we will be considering are the averages
 and variances of the oscillator coordinate and momentum, as well as 
$Tr\rho^2$ as a measure of oscillator entanglement to the bath.
We also would like to show how different initial states for the bath modes 
lead to different exact master equations for the oscillator density matrix.    
In deriving such equations we will use an exact formal solution for the 
characteristic function of the oscillator, rather then the path integral
techniques for the reduced density matrix \cite{Hu,Hu2,Romero,Grabert}. 
This approach makes it possible for this model to obtain 
master equations for arbitrary factorized initial conditions.

This paper is organized as follows. In Sec. \ref{oscil} we
consider the model and its exact solution. In Sections \ref{equili}, 
\ref{numbersta}
and \ref{cohsta} 
the oscillator behavior is considered, respectively, for the bath modes 
in equilibrium, number states and coherent states.
Exact master equations are discussed in \ref{equations}. 
Concluding remarks are given in Sec. \ref{conc}


\section{The model of a linearly coupled oscillator} \label{oscil}
The system Hamiltonian is given by
\bea
H&=&\hbar\nu a^{\dagger}a + \sum_{k}\hbar\om_kb_k^{\dagger}b_k  
+\sum_k\hbar u_ka^{\dagger}b_k \non \\ 
& &+ \sum_k\hbar
u^{*}_kb_k^{\dagger}a 
+\sum_k\hbar v_ka^{\dagger}b_k^{\dagger}+ \sum_k\hbar
v_k^{*}b_ka.\label{ham}
\eea
Here $a^{\dag}$ and $a$ are the creation and annihilation operators of
the harmonic oscillator, and $b^{\dagger}_k$ and $b_k$ are the
creation and annihilation operators for the bath modes.  
The coefficients are assumed to be such that the Hamiltonian is a positive 
definite quadratic form. 
 The coordinate and momentum operators for the oscillator are related
to $a^{\dag}$ and $a$ through
\be
x=\sqrt{\frac{\hbar}{2m\nu}}(a^{\dag} + a),\qquad
p=i\sqrt{\frac{\hbar m\nu}{2}}(a^{\dag} - a),
\ee
where $m$ is the oscillator mass.
By a suitable choice of coefficients Hamiltonian (\ref{ham}) reduces
to the Hamiltonian with coordinate-coordinate coupling
or the rotating wave approximation (RWA) Hamiltonian.
In particular, in the latter case the last two terms
in (\ref{ham}) are dropped.
We will assume that if the number of the bath modes increases to infinity
the  frequency $\om_k$ becomes a continuous function of $k$.
We will refer to the limit of the infinite number of modes with average
energy of each mode being held constant as the thermodynamic limit.

Various forms of Hamiltonian (\ref{ham}) corresponding to different choices
of frequencies and coupling constants as well as its exact diagonalization
has been extensively studied in the literature. General but formal discussion
of the diagonalization of a Hermitian quadratic  bosonic form [of which 
(\ref{ham}) is a special case] can be found in \cite{Bogoliubov}. Systems of 
 oscillators with
coordinate only coupling were considered in \cite{Ford}. 
Detailed investigation of Hamiltonian (\ref{ham}) in the case of coordinate
coupling can be found in references \cite{Haake,Hakim,Katja}.
Relation between several forms of (\ref{ham}) and the translationally 
invariant Hamiltonian with coordinate coupling was discussed in \cite{Ford2}.
Relation between the coordinate coupling and RWA 
is studied in \cite{Katja,daCosta}. 
 
The equations of motion for the annihilation and creation operators are 
\bea
& &\dot{a}=-i\nu a -i\sum_ku_kb_k-i\sum_kv_kb_k^{\dag}, \non \\
& &\dot{b}_k=-i\om_kb_k-iu^{*}_ka-iv_ka^{\dag},  \non \\
& &\dot{a^{\dag}}=i\nu a^{\dag} +i\sum_ku^{*}_kb_k^{\dag}+i\sum_kv_k^{*}b_k, 
\non \\
& &\dot{b}_k^{\dag}=i\om_kb_k^{\dag}+iu_ka^{\dag}+iv^{*}_ka.
\label{linsys}
\eea
This system of equations can be solved subject to the set of initial 
 conditions
$
a(0)=a$, $b_k(0)=b_k$, $a^{\dag}(0)=a^{\dag}$, and $b_k^{\dag}(0)=b^{\dag}_k$.
Since system (\ref{linsys}) is linear, its solutions will depend linearly 
on the initial conditions. In particular, $a(t)$ is given by
\be
a(t)=A(t)a +\sum_kB_k(t)b_k+C(t)a^{\dag} +\sum_kD_k(t)b_k^{\dag}. \label{a}
\ee
Similarly, for $a^{\dag}(t)$ we have
\be
a^{\dag}(t)=A^*(t)a^{\dag} +\sum_kB^*_k(t)b^{\dag}_k+ C^*(t)a 
+\sum_kD^*_k(t)b_k\label{adag}.
\ee
Coefficients $A(t)$, $B_k(t)$, $C(t)$, and $D_k(t)$ satisfy the following 
relation (see \cite{Bogoliubov} for details):
\be
\bigl|A(t)\bigr|^2-\bigl|C(t)\bigr|^2
+\sum_k\bigl|B_k(t)\bigr|^2-\sum_k\bigl|D_k(t)\bigr|^2=1.
\ee
They can, in principle,
be calculated for a each particular form of frequencies and coupling 
parameters. For the purposes of this paper we will not need explicit 
expressions for these coefficients.
The only assumptions we will use is that 
in the thermodynamic limit coefficients $A(t)$ and $C(t)$ vanish 
for $t\to\infty$,  and $B_k(t)$ and $D_k(t)$ remain bounded in the same 
limit. Physically these requirements correspond to the fact the the
initial state of the oscillator is forgotten for long times while
any observables associated with it (e.g., average energy)
remain finite. 
The detailed calculations of coefficients that show such behavior as
well as conditions on the coupling constants and frequencies in the 
thermodynamic limit can be found
in the original references \cite{Caldeira,Hakim,Haake,Katja,Grabert}.

The reduced dynamics of the oscillator is conveniently  described in terms 
of a symmetrically ordered characteristic function defined as \cite{Walls} 
\be 
\chi(\eta,t)=Tr\bigl(\rho e^{\eta a^{\dag}-\eta^{*}a}\bigr).\label{char}
\ee  
Here $\chi(\eta,t)$ can be treated as a function of $\eta$, $\eta^*$ 
and $t$.
We will always suppress the second variable to simplify the notation.
Using characteristic function (\ref{char}) we can calculate expectation 
values of the symmetrized products of operators $a^{\dag}$ and $a$. 
It can also be converted to any of the quasi-probability 
distribution functions or the reduced density matrix \cite{Walls}.
We will also use $Tr \rho^2$ as a measure of purity of the oscillator 
state $\rho$. In terms of the characteristic function $Tr \rho^2$ is given 
by
\be
Tr \rho^2=\frac{1}{\pi}\int d^2\eta \bigl| \chi(\eta) \bigr|^2. \label{trace}
\ee

To see how $\chi(\eta,t)$ evolves in time we use the 
Heisenberg picture and insert expressions (\ref{a}) 
and (\ref{adag}) into (\ref{char}) 
to obtain 
\begin{widetext}
\be
\chi(\eta,t)
=Tr\left(\rho e^{(\eta A^{*}-\eta^{*} C)a^{\dag}-(\eta^{*} A-\eta C^{*})a}
\prod_k e^{(\eta B_k^{*}-\eta^{*} D_k)b_k^{\dag}
-(\eta^{*} B_k-\eta D_k^{*})b_k}\right).
\ee 
\end{widetext}
The time dependence of the coefficients is suppressed here and whenever 
possible to avoid heavy notation.

In this paper we consider only factorized initial conditions, i.e., we 
assume that the initial density matrix  of the whole system 
 $\rho_{total}$  factorizes into the oscillator and bath density matrices 
 as $\rho_{total}=\rho\otimes\rho_{bath}$. In this case the characteristic 
function takes the form
\be 
\chi(\eta,t)=\chi\left((\eta A^{*}-\eta^{*} C),0\right)F(\eta,t), 
\label{solu1}
\ee 
with
\be
F(\eta,t)
= Tr\left(\rho_{bath}\prod_k e^{(\eta B_k^{*}-\eta^{*} D_k)b_k^{\dag}
-(\eta^{*} B_k-\eta D_k^{*})b_k}\right).
\ee
Equation (\ref{solu1}) expresses the  oscillator characteristic function 
at time $t$ in terms of the initial characteristic function. In the limit 
of long times, when $A$ and $C$ vanish, characteristic function  
 $\chi(\eta,t)$ is determined by the asymptotic form of $F(\eta,t)$.

We will now consider the time evolution and asymptotic values 
of the average oscillator 
coordinate and momentum, the coordinate and momentum variances and $Tr\rho^2$ 
for different initial states of the bath. 

\section{Equilibrium state of the bath} \label{equili}
This section consists primarily of an overview of well-known results.
The state of the bath is the grand canonical ensemble given by 
\be
\rho_{eq}=\prod_k\bigl(1-e^{-\beta\hbar\om_k}\bigr)
e^{-\beta\hbar\om_kb_k^{\dagger}b_k }\label{Gibbs}.
\ee
Using the identity
\be
Tr\left(\rho_{eq}e^{\eta b_k^{\dag}-\eta^* b_k}\right)
=e^{-|\eta|^2(\overline{n}_k+\frac{1}{2})},
\ee
with 
\be
\overline{ n}_k=\frac{1}{e^{\beta\hbar\om_k}-1},\label{ocnum}
\ee
 we find 
function $F(\eta,t)$ to be
\be
F(\eta,t)
=e^{-\alpha | \eta |^2+\gamma^{*}\eta^2+\gamma {\eta^{*}}^2}. 
\label{eqchar} 
\ee  
Here coefficients $\alpha$ and $\gamma$ are given by
\bea
\alpha&=&\sum_k\left(\bigl| B_k \bigr|^2+\bigl |D_k\bigr|^2\right)
\left(\overline {n}_k+\frac{1}{2}\right), \non \\ 
\gamma&=&\sum_kB_kD_k\left(\overline{n}_k+\frac{1}{2}\right). 
\eea 
We will often use coefficients $\alpha$ and $\gamma$ in the limits 
$t\to\infty$ or $T\to0$. In such cases we will use the notation 
$\alpha_{\infty}$, $\gamma_{\infty}$ and $\alpha^0$, $\gamma^0$, 
respectively. If both limits are taken we will use $\alpha^0_{\infty}$ 
and $\gamma^0_{\infty}$.

The time evolution of the average coordinate and momentum is easily 
calculated either through the characteristic function, or by directly 
using (\ref{a}) and (\ref{adag}).
\bea
\bigl\bra x(t)\bigr\ket
&=&\sqrt{\frac{\hbar}{2m\nu}}\bigl((A^*+C)\bra a^{\dag}\ket
+(A+C^*)\bra a\ket\bigr),\non \\
\bigl\bra p(t)\bigr\ket
&=&i\sqrt{\frac{\hbar m\nu}{2}}\bigl((A^*+C)\bra a^{\dag}\ket-
(A+C^*)\bra a\ket
\bigr).\label{aveq}
\eea
There is no dependence on the state of the bath. 
For arbitrary temperature $\bra x(t)\ket$ and $\bra p(t)\ket$ depend only 
on their initial average values. In the limit of infinitely long times 
 $\bra x(t)\ket$ and $\bra p(t)\ket$ vanish.

For the variances of the oscillator coordinate and momentum we obtain
\begin{widetext}
\bea
\bigl\bra x^ 2(t)\bigr\ket -\bigl\bra x(t)\bigr\ket ^2
&=&\frac{\hbar}{2m\nu}\Bigl[\bigl(A^*+C\bigr)^2
\bigl(\bra a^{\dag}a^{\dag}\ket-
\bra a^{\dag}\ket \bra a^{\dag}\ket\bigr)+\bigl(A+C^*\bigr)^2
\bigl(\bra aa\ket
-\bra a\ket \bra a\ket\bigr)
\non \\
& &+2\bigl|\bigl(A+C^*\bigr)\bigr|^2\bigl(\bra a^{\dag}a\ket
-\bra a^{\dag}\ket \bra a\ket+\frac{1}{2}\bigr)
+2(\alpha+\gamma+\gamma^*)\Bigr],  \non \\
\bigl\bra p^ 2(t)\bigr\ket -\bigl\bra p(t)\bigr\ket ^2
&=& \frac{\hbar m\nu}{2}\Bigl[-\bigl(A^*-C\bigr)^2
\bigl(\bra a^{\dag}a^{\dag}\ket-
\bra a^{\dag}\ket \bra a^{\dag}\ket\bigr)-\bigl(A-C^*\bigr)^2
\bigl(\bra aa\ket
-\bra a\ket \bra a\ket\bigr) \non \\
 & &+2\bigl|\bigl(A-C^*\bigr)\bigr|^2\bigl(\bra a^{\dag}a\ket
-\bra a^{\dag}\ket \bra a\ket+\frac{1}{2}\bigr)
+2(\alpha-\gamma-\gamma^*)\Bigr]\label{pvar}.
\eea
\end{widetext}
These quantities depend both on the initial state of the oscillator 
and the temperature of the bath. 
For $t\to\infty$ we have
\bea
\bra x^ 2\ket -\bra x\ket ^2
&=&\frac{\hbar}{m\nu}(\alpha_{\infty}
+\gamma_{\infty}+\gamma^*_{\infty}), \non \\
\bra p^ 2\ket -\bra p\ket ^2
&=&{\hbar m\nu}(\alpha_{\infty}-\gamma_{\infty}-\gamma^*_{\infty}).
\label{pvarasy}
\eea
Coefficients $\alpha_{\infty}$ and $\gamma_{\infty}$ are 
proportional to $\overline{n}_k$.
Therefore, at high temperatures variances (\ref{pvarasy}) grow as $kT$.

The measure of the oscillator purity is given by the integral:
\be
Tr\rho^2= 
\frac{1}{\pi}\int\! d^2\eta|\chi((\eta A^{*}-\eta^{*} C),0)|^2
e^{-2(\alpha | \eta |^2-\gamma^*\eta^2-\gamma {\eta^{*}}^2)}.
\label{eqtrace} 
\ee
This integral can be calculated for specific initial states of 
the oscillator.
For infinitely long times, when the initial state is forgotten, 
we obtain
\be
Tr\rho^2=
\frac{1}{2\sqrt{\alpha^2_{\infty}-4|\gamma_{\infty}|^2}}.
\ee
For high temperatures $Tr\rho^2$ is proportional to $1/kT$. 
The state of the oscillator 
becomes less pure as the temperature grows. 
Let us note that for intermediate times $Tr\rho^2$ can  take  lower 
values than its value at $t\to\infty$. 
For zero temperature of the bath the asymptotic value of $Tr\rho^2$ 
will, in general, be less then one since the oscillator remains 
dressed at $T=0$. 

For the special case of the RWA Hamiltonian, coefficients $C(t)$ and 
$D_k(t)$ in (\ref{a}) and (\ref{adag})
are equal to zero. As a result, 
$\alpha^0_{\infty}=\frac{1}{2}$ and
$\gamma=0$ and for long times $Tr\rho^2=1$.
In this case the reduced vacuum of the oscillator is a pure state that 
is identical to the ground state of the uncoupled oscillator.

\section{The number states for the bath modes}\label{numbersta}
We now consider the initial state of the bath with each mode in a number 
state. 
\be
\bigl|\{n_k\}\bigr\ket
=\bigl|n_{k_1}\bigr\ket\otimes\bigl|n_{k_2}\bigr\ket\otimes\cdots
\ee
with
$\{n_k\}$ denoting a set of occupation numbers $n_k$ for all modes.
 
Function $F(\eta,t)$ can be calculated using the identity \cite{Glauber10}
\be
\bra n_k| e^{\eta b_k^{\dag}-\eta^{*}b_k}| n_k \ket
=e^{-\frac{|\eta |^2}{2}}L_{n_k}\bigl(|\eta |^2\bigr),
\ee
where $L_{n}(x)$ is a Laguerre polynomial.
We obtain
\bea
F(\eta,t)&=&e^{-\alpha^0 | \eta |^2
+{\gamma^0}^*\eta^2+\gamma^0 {\eta^*}^2} \non \\
& &\times\prod_k L_{n_k}
\bigl(\left| (\eta {B^*}_k-\eta^{*} D_k)\right|^2\bigr). \label{numchar}
\eea

The behavior of average coordinate and momentum of the oscillator 
is exactly the same as for the equilibrium state of the bath and 
 given by (\ref{aveq}). 
Thus,  $\bra x(t)\ket$ and $\bra p(t)\ket$
do not depend on the particular set of occupation numbers. Both quantities
 vanish for $t\to\infty$.
For the variances of the coordinate and momentum we have
\begin{widetext}
\bea
\bigl\bra x^ 2(t)\bigr\ket -\bigl\bra x(t)\bigr\ket ^2
&=&\frac{\hbar}{2m\nu}\Bigl[\bigl(A^*+C\bigr)^2
\bigl(\bra a^{\dag}a^{\dag}\ket-
\bra a^{\dag}\ket \bra a^{\dag}\ket\bigr)+\bigl(A+C^*\bigr)^2
\bigl(\bra aa\ket
-\bra a\ket \bra a\ket\bigr) \non \\
& &+2\bigl|\bigl(A+C^*\bigr)\bigr|^2\bigl(\bra a^{\dag}a\ket
-\bra a^{\dag}\ket \bra a\ket+\frac{1}{2}\bigr)
+2(\tilde{\alpha}+\tilde{\gamma}+\tilde{\gamma}^*)\Bigr], \non \\
\bigl\bra p^ 2(t)\bigr\ket -\bigl\bra p(t)\bigr\ket ^2
&=& \frac{\hbar m\nu}{2}\Bigl[-\bigl(A^*-C\bigr)^2
\bigl(\bra a^{\dag}a^{\dag}\ket-
\bra a^{\dag}\ket \bra a^{\dag}\ket\bigr)-\bigl(A-C^*\bigr)^2
\bigl(\bra aa\ket
-\bra a\ket \bra a\ket\bigr) \non \\
 & &+2\bigl|(A-C^*)\bigr|^2\bigl(\bra a^{\dag}a\ket
-\bra a^{\dag}\ket \bra a\ket+\frac{1}{2}\bigr)
+2(\tilde\alpha-\tilde\gamma-\tilde\gamma^*)\Bigr]\label{pnumvar},
\eea
\end{widetext}
where
\bea
\tilde\alpha&=&
\sum_k\left(\bigl| B_k \bigr|^2+\bigl |D_k\bigr|^2\right)
\left(n_k+\frac{1}{2}\right), \non \\
\tilde\gamma&=&
\sum_kB_kD_k(n_k+\frac{1}{2}).
\eea
The variances differ from the equilibrium ensemble case by
the replacement of $\overline n_k $ with  $n_k$.

To get a better   
picture of how these values relate to the equilibrium case 
we have to make some assumption about the occupation numbers.  
Let  us consider an ensemble of the occupation number states 
corresponding to the number state  decomposition of 
the equilibrium density matrix
\be
\rho_{eq}=\sum_{\{ n_k \} }\bigl|\{n_k\}\bigr\ket
P_n\bigl(\{n_k\}\bigr)\bigl\bra \{n_k\}\bigr| \label{eirez}.
\ee
Here the probability for a particular set of occupation numbers 
$P_n\bigl(\{n_k\}\bigr)$ is given by
\be
P_n\bigl(\{n_k\}\bigr)
=\prod_k\bigl(1-e^{-\beta\hbar \om_k}\bigr)e^{-\beta\hbar\om_kn_k}.
\ee 
We now assume that the number states are taken from  
ensemble (\ref{eirez}). Any quantities calculated for each individual
number state (e.g., averages, variances, $Tr\rho^2$) can then be 
averaged over $P_n\bigl(\{n_k\}\bigr)$ to obtain their average values 
in ensemble (\ref{eirez}). These latter averages will give typical
values for the pure state quantities in ensemble (\ref{eirez}).

Averaging  variances (\ref{pnumvar}) over ensemble (\ref{eirez}) 
will give the same variances as for the equilibrium case. In particular,
in the limit of long times we have
\bea
\overline{\bigl\bra x^ 2\bigr\ket -\bigl\bra x\bigr\ket ^2}
&=&\frac{\hbar}{m\nu}(\alpha_{\infty}
+\gamma_{\infty}+\gamma^*_{\infty}), \non \\ 
\overline{\bigl\bra p^ 2\bigr\ket -\bigl\bra p\bigr\ket ^2}
&=&{\hbar m\nu}(\alpha_{\infty}
-\gamma_{\infty}-\gamma^*_{\infty}).\label{pnumvarasy}
\eea
Here we use overlining to denote averaging over the ensemble of pure 
states.

The fact that the average variances are the same as 
for the equilibrium ensemble is not surprising since expressions in 
(\ref{pnumvar}) are linear in $n_k$'s.
More importantly, in the thermodynamic limit almost all states 
in ensemble (\ref{eirez}) will have the same variances as in equilibrium. 
Let us show this for the coordinate variance.  
We can treat the coordinate variance as a function of 
random variables $n_k$ described by the distribution  
$P_n\bigl(\{n_k\}\bigr)$. Calculating the variance of this 
function for the distribution $P_n\bigl(\{n_k\}\bigr)$ we obtain.
\bea
& &\overline{\bigl(\bra x^2\ket -\bra x\ket ^2\bigr)^2}-
\overline{\bigl(\bra x^2\ket -\bra x\ket ^2\bigr)}
\cdot\overline{\bigl(\bra x^2\ket -\bra x\ket ^2\bigr)}\non \\
& &=
\frac{\hbar^2}{m^2{\nu}^2}\sum_k\bigl|\bigl(B_k+D^*_k\bigr)\bigr|^4
\left(\overline{n_k^2}-{\overline{n}_k}^2\right). \label{variance}
\eea
We note that coefficients $B_k$ and $D_k$ must depend on the number of 
bath modes $N$ as
$1/\sqrt N$ in order for the quantities like (\ref{pnumvar}) to remain 
finite in the thermodynamic limit. Therefore the sum over $k$ 
in (\ref{variance}) is proportional to $1/N$ and vanishes for $N\to \infty$. 
A similar argument can be applied to
$ \bigl(\bra p^2(t)\ket -\bra p(t)\ket ^2\bigr)$.
Thus, as in equilibrium case, we expect the variances to grow as $kT$ 
for high occupation numbers.  

Let us now consider the behavior of $Tr\rho^2$. 
Using definition (\ref{trace}) and characteristic function (\ref{numchar}) 
we obtain 
\begin{widetext}
\be
Tr\rho^2=
\frac{1}{\pi}\int d^2\eta
\bigl|\chi\left((\eta A^{*}-\eta^{*} C\right),0)\bigr|^2
e^{-2(\alpha^0 | \eta |^2-{\gamma^0}^*\eta^2-\gamma^0{\eta^*}^2)}
\prod_k{L^2_{n_k}}\left(\bigl| (\eta {B^*}_k-\eta^{*} D_k)\bigr|^2\right).
\ee
\end{widetext}
If number states are taken from the ensemble (\ref{eirez}) we can 
calculate average $Tr\rho^2$ for number states in this ensemble. 
Using the identity \cite{Gradshtein}
\be
\sum_nL^2_n(x)z^n=
\frac{1}{1-z}\exp{\Bigl( -\frac{2zx}{1-z}\Bigr)},\qquad \bigl(|z|<1\bigr),
\ee
we obtain
\be
\overline{Tr\rho^2}=
\frac{1}{\pi}\int\!d^2\eta|\chi((\eta A^{*}-\eta^{*} C),0)|^2
e^{-2(\alpha | \eta |^2-\gamma^{*}\eta^2-\gamma {\eta^{*}}^2)}.
\ee
This is exactly the same as $Tr\rho^2$ in (\ref{eqtrace}).
We can see that, at least on the average,  $Tr\rho^2$ for number states 
from ensemble (\ref{eirez}) is the same at all times as for the case of 
 equilibrium bath. As a consequence, in the limit of long times the state 
 of the oscillator becomes less pure for higher occupation numbers for
 the bath modes.

\section{Coherent states for the bath modes} \label{cohsta}
We now consider the case where all bath modes are initially in coherent 
states. 
\be
\bigl|\{\beta_k\}\bigr\ket
=\bigl|\beta_{k_1}\bigr\ket\otimes\bigl|\beta_{k_2}\bigr\ket\otimes\cdots
\ee
with $\{\beta_k\}$ denoting a set of complex numbers $\beta_k$ 
specifying the coherent
states.
One can 
interpret such a state as the most classical state of the bath. 
Function $F(\eta,t)$ is  calculated to be 
\be
F(\eta,t)=e^{\delta^*\eta-\delta\eta^*-\alpha^0 | \eta |^2
+{\gamma^0}^*\eta^2+\gamma^0 {\eta^{*}}^2}. \label{cohchar}
\ee
Here 
\be
\delta=\sum_k\bigl(B_k\beta_k+D_k\beta^*_k\bigr). \label{delat}
\ee

For the average coordinate and momentum we obtain
\bea
\bigl\bra x(t)\bigr\ket&=&
\sqrt{\frac{\hbar}{2m\nu}}
\bigl((A^*+C)\bra a^{\dag}\ket+(A+C^*)\bra a\ket \non  \\ 
& &+\delta+\delta^*\bigr), \non \\ 
\bigl\bra p(t)\bigr\ket&=&
i\sqrt{\frac{\hbar m\nu}{2}}\bigl((A^*-C)\bra a^{\dag}\ket-
(A-C^*)\bra a\ket \non \\ 
& &+\delta^*-\delta\bigr).\label{avco}
\eea
To get a better understanding  of how $\bra x(t)\ket$ and 
$\bra p(t)\ket$ behave we have to make some assumption about 
parameters $\beta_k$.

Analogously to the procedure used for the number state bath 
let us assume that 
the coherent states are taken from the ensemble 
corresponding to the coherent states decomposition of the equilibrium 
density matrix
\be
\rho_{eq}=
\int d^2\{\beta_k\}\bigl|\{\beta_k\}\bigr\ket
P_c\bigl(\{\beta_k\}\bigr)\bigl\bra\{\beta_k \}\bigr|.\label{coh}
\ee
The probability distribution $P_c\bigl(\{\beta_k\}\bigr)$ is given by
\be
P_c\bigl(\{\beta_k\}\bigr)=\prod_k
\frac{1}{\pi\overline {n}_k}
\exp\left({\frac{-|\beta_k|^2}{\overline {n}_k}}\right). \label{cohprob}
\ee
Decomposition (\ref{coh}) is just the $P$-representation of 
the equilibrium density matrix \cite{Walls}.

If the coherent states are taken from ensemble (\ref{coh}), then,
for each set of the coherent states, $\delta(t)$ is a realization 
of a complex colored normal noise with
zero mean and correlation functions given by
\bea
\overline{\delta(t)\delta^*(t')}
=\sum_k\left(B_k(t)B^*_k(t')+D_k(t)D^*_k(t')\right)\overline n_k, \non \\ 
\overline{\delta(t)\delta(t')}
=\sum_k\left(B_k(t)D_k(t')+B_k(t')D_k(t)\right)\overline n_k. 
\label{corred}
\eea
Clearly,
$\bra x(t)\ket$ and $\bra p(t)\ket$ are  also 
realizations of the normal noise. The mean in this case is, in general,
non-zero but will go to zero for long times.

Let us evaluate the typical asymptotic values taken by $\bra x(t)\ket$ 
and $\bra p(t)\ket$. This can be done by calculating average
${\bra x\ket}^2$ and ${\bra p\ket}^2$ for ensemble (\ref{coh}). 
\be
\overline{{\bra x\ket}^2}
=\frac{\hbar}{m\nu}\left(\alpha_{\infty}+\gamma_{\infty}
+\gamma^*_{\infty}-\alpha^0_{\infty}-\gamma^0_{\infty}
-{{\gamma^0}_{\infty}}^*\right),
\ee
\be
\overline{{\bra p \ket}^2}
={\hbar m\nu}\left(\alpha_{\infty}-\gamma_{\infty}
-\gamma^*_{\infty}-\alpha^0_{\infty}+\gamma^0_{\infty}
+{{\gamma^0}_{\infty}}^*\right).  \label{psqco}
\ee
For high temperatures these quantities are proportional to $kT$. 
Therefore typical asymptotic values for $\bra x\ket$ and $\bra p\ket$ 
are proportional to $\sqrt{kT}$. 

The coordinate and momentum variances are the same as for the 
case of equilibrium bath at zero temperature, i.e., they are given
by (\ref{pvar}) with $\alpha$ and $\gamma$ replaced by
$\alpha^0$ and $\gamma^0$, respectively.
Therefore, these variances do not depend on a particular set of 
parameters $\beta_k$ and  their asymptotic values are the same as 
for the reduced vacuum state of the oscillator. 

The value of $Tr\rho^2$ is given by
\begin{widetext}
\be
Tr\rho^2=
\frac{1}{\pi}\int\! d^2\eta
\left|\chi\left((\eta A^{*}-\eta^{*} C),0\right)\right|^2
e^{-2(\alpha_0 | \eta |^2-\gamma_0^*\eta^2-\gamma_0 {\eta^{*}}^2)},
\label{encoh}
\ee
\end{widetext}
and is the same as in the case of zero temperature bath with the 
asymptotic value given by
\be
Tr\rho^2=
\frac{1}{2\sqrt{{\left(\alpha^0_{\infty}\right)}^2
-4\bigl|{\gamma^0_{\infty}}\bigr|^2}}.
\ee

More generally, the asymptotic characteristic
function is determined by the asymptotic form of function $F(\eta,t)$ 
(\ref{cohchar}) and corresponds to the density matrix that is 
a displaced reduced vacuum state $\rho_{vac}$,
\be
\rho=D(\delta)\rho_{vac}D^{\dag}(\delta),
\ee
where the displacement operator $D(\delta)$ is given by
\be
D(\delta)=e^{\delta a^{\dag}-\delta^*a}.
\ee
 
We can conclude that for coherent states bath in the limit of long times
the oscillator 
becomes localized in the following sense: the coordinate and momentum 
variances, as well as entanglement to the bath, are same as for 
the reduced vacuum (and do not depend on temperature), 
while the average coordinate and momentum randomly fluctuate with 
the amplitude of the fluctuations proportional to
$\sqrt{kT}$. The rate of such localization is of the order of 
the relaxation rate as can be concluded from (\ref{pvar}) for $T=0$.
 
For the special case of the RWA Hamiltonian the asymptotic state of 
the oscillator is a pure coherent state. The oscillator can start 
in a mixed or pure state, 
it then goes through a period of entanglement and asymptotically its 
state becomes a pure coherent state with randomly fluctuating 
average coordinate and momentum. If the oscillator is initially in 
a coherent state it will always remain in a coherent state, 
never entangling with the bath \cite{Glauber3}.
In Appendix we give explicit expression for $Tr\rho^2(t)$ for two
types of initial oscillator states for the RWA case.
 
\section{Exact master equations} \label{equations}
There is a similarity of a mathematical nature between the coherent states 
bath and bath in equilibrium. In both cases function $F(\eta,t)$ is 
Gaussian, and, as a result, the characteristic function satisfies 
simple master equations as will be shown shortly. 

Before doing so let us note that for any factorized initial conditions 
(for any system-bath model) there always exists an exact equation for 
the reduced density matrix with the time evolution governed by a 
time-dependent operator that does not depend on the state of the system. 
Obtaining an explicit expression for such an operator can, in general, 
be a difficult task. Let us show that such an operator can be constructed 
for the present model. We will continue to use the characteristic function 
space because of its mathematical convenience. If necessary the equations 
can be transformed into other representations.
To simplify the demonstration we introduce the following
transformed characteristic function
\be
 \tilde{\chi}(\eta,t)=\frac{\chi(\eta,t)}{F(\eta,t)}.\label{tran}
\ee
Using expression (\ref{solu1}) for $\chi(\eta,t)$ and the fact 
that $\tilde{\chi}(\eta,0)=\chi(\eta,0)$ we can write 
function $\tilde{\chi}(\eta,t)$ at time $t$ in terms of 
the $\tilde{\chi}(\eta,0)$ as 
\be
\tilde{\chi}(\eta,t)=\tilde{\chi}\left((\eta A^*-\eta^*C),0\right). 
\label{transolu}
\ee
We now use the fact that $\tilde{\chi}(\eta,t)$ depends on $\eta$,
$\eta^*$ and $t$ only through $(\eta A^*-\eta^*C)$ 
and $(\eta^* A-\eta C^*)$.
Differentiating  solution (\ref{transolu}) with respect to time we have
\bea
\frac{\partial \tilde{\chi}}{\partial t}&=&\frac{\partial \tilde{\chi}}
{\partial(\eta A^*-\eta^*C)}(\eta \dot{A^*}-\eta^*\dot{C}) \non \\ 
& &+\frac{\partial \tilde{\chi}}
{\partial(\eta^* A-\eta C^*)}(\eta^* \dot{A}-\eta\dot{C^*}). \label{ptime}
\eea
For derivatives with respect to $\eta$ and $\eta^*$ we obtain
\bea
\frac{\partial\tilde{\chi}}{\partial \eta}=
\frac{\partial \tilde{\chi}}{\partial(\eta A^*-\eta^*C)}A^*
-\frac{\partial \tilde{\chi}}{\partial(\eta^* A-\eta C^*)}C^*, \non \\ 
\frac{\partial\tilde{\chi}}{\partial \eta^*}=
-\frac{\partial \tilde{\chi}}{\partial(\eta A^*-\eta^*C)}C
+\frac{\partial \tilde{\chi}}{\partial(\eta^* A-\eta C^*)}A.
\eea 
Solving the last two equations for the derivatives of $\tilde{\chi}$ with
respect to $(\eta A^*-\eta^*C)$ and $(\eta^* A-\eta C^*)$ and substituting 
them in (\ref{ptime}) we obtain a closed equation for $\tilde{\chi}$:
\be
\frac{\partial\tilde{\chi}}{\partial t}
=\left(\xi^*(t)\eta+\zeta(t)\eta^*\right)
\frac{\partial \tilde{\chi}}{\partial \eta}
+\left(\xi(t)\eta^*+\zeta^*(t)\eta\right)
\frac{\partial \tilde{\chi}}{\partial \eta^*}.
\label{traneq}
\ee
Here $\xi(t)$ and $\zeta(t)$ are given by
\be
\xi(t)=\frac{A^*\dot{A}-C^*\dot{C}}{|A|^2-|C|^2}
,\qquad \zeta(t)=\frac{C\dot{A}-A\dot{C}}{|A|^2-|C|^2}.
\ee
Substituting definition of $\tilde{\chi}$ (\ref{tran}) into 
equation (\ref{traneq}) 
we obtain the following equation for ${\chi}$ 
\begin{widetext}
\bea
\frac{\partial {\chi}}{\partial t}
&=&\left(\xi^*(t)\eta+\zeta(t)\eta^*\right)
\left(\frac{\partial {\chi}}{\partial \eta}
-\left(\frac{\partial\ln F}{\partial\eta}\right)\chi\right)
+\left(\xi(t)\eta^*+\zeta^*(t)\eta\right)
\left(\frac{\partial {\chi}}{\partial \eta^*}
-\left(\frac{\partial\ln F}{\partial\eta^*}\right)\chi\right)\non \\
& &+\left(\frac{\partial\ln F}{\partial t}\right)\chi. \label{theequation}
\eea
\end{widetext}
This is a closed equation for characteristic function $\chi(\eta,t)$.
The explicit form of the time-dependent operator is determined 
by function $F(\eta,t)$, which, in turn, is determined by the initial 
state of the bath.
We will not go into the analysis of this equation and limit ourselves 
to a few remarks. This equation is always local in the 
$\eta$-space. However, transforming it into the quasi-probability 
distribution space or coordinate or momentum representation for the 
density matrix will, 
in general, lead to non-local equations. Only for some special forms 
of $F(\eta,t)$ can we expect to get local
 equations in these representations.  Note that 
 exact master equations for the oscillator-bath model
that appeared in the literature use equilibrium initial state
for the bath \cite{Haake,Hu}, squeezed equilibrium for the bath \cite{Hu2},
 or modified equilibrium for the whole
system \cite{Romero,Grabert}. When converted into the characteristic
function space all these equations contain time dependent operators
that are at most quadratic in $\eta$, $\eta^*$, 
$\partial/\partial\eta$, and $\partial/\partial\eta^*$. This form
of the time dependent operator is related to the fact that 
propagating function for such initial conditions is Gaussian. 
Inspection of equation (\ref{theequation}) shows that such a 
simple form of the time dependent operator in our case is possible 
only for Gaussian functions $F(\eta,t)$.
 
We now consider particular form of equation (\ref{theequation}) for 
this special case, i.e., when function $F(\eta,t)$ is given by
\be
F(\eta,t)
=e^{{\underline\delta}^*\eta-{\underline\delta}\eta^*
-{\underline\alpha} |\eta |^2+{\underline\gamma}^*\eta^2
+\underline{\gamma} {\eta^{*}}^2}.
\label{GaussF}
\ee
Here $\underline\delta$, $\underline\alpha$ and $\underline\gamma$ are 
time dependent parameters characterizing each
particular Gaussian state of the bath.
Function (\ref{GaussF}) 
will include such states for the bath modes as equilibrium, squeezed
states, squeezed and displaced equilibrium, etc.
Substituting (\ref{GaussF}) into (\ref{theequation}) we obtain 
after some simplifications
\bea
\frac{\partial {\chi}}{\partial t}&=&
\left(\xi^*(t)\eta+\zeta(t)\eta^*\right)
\frac{\partial {\chi}}{\partial \eta}
+\left(\xi(t)\eta^*+\zeta^*(t)\eta\right)
\frac{\partial {\chi}}{\partial \eta^*} \non \\
& &+\kappa(t)|\eta|^2\chi+\mu^*(t)\eta^2\chi
+\mu(t){\eta^*}^2\chi \non \\
& &+\sigma^*(t)\eta\chi-\sigma(t)\eta^*\chi. \label{gaueq}
\eea
Coefficients $\kappa(t)$, $\mu(t)$ and $\sigma(t)$ are given by
\bea
\kappa(t)&=&{\underline\alpha}(\xi+\xi^*)
-2(\zeta{\underline\gamma}^*+\zeta^*{\underline\gamma})
-\dot{{\underline\alpha}}, \non \\ 
\mu(t)&=& \zeta{\underline\alpha}-2\xi{\underline\gamma}
+\dot{{\underline\gamma}}, \non \\ 
\sigma(t)&=&\zeta{\underline\delta}^*-\xi{\underline\delta}
+\dot{{\underline\delta}}. \label{coeffi}
\eea

In the case of the equilibrium bath we need to put 
$\underline\alpha=\alpha$, $\underline\gamma=\gamma$ and 
$\underline\delta=0$ in (\ref{coeffi}). 
In this case the last
two terms on the right-hand side of equation (\ref{gaueq}) disappear and
the equation becomes the characteristic function version of the 
Hu-Paz-Zhang equation \cite{Hu,Haake} for the Hamiltonian (\ref{ham}),
 i. e., it will allow for the possibility of momentum-momentum and 
momentum-coordinate coupling between the bath and the oscillator. 

If the bath modes are initially in coherent states then we have to use
$\underline\alpha=\alpha^0$,
$\underline\gamma=\gamma^0$ and $\underline\delta=\delta$.
If the coherent states are taken from ensemble (\ref{coh}), then, for each 
particular set of states, $\sigma(t)$
will be a realization of a complex colored normal noise. The equation
becomes a stochastic master equation with a multiplicative noise. 
In this case (\ref{gaueq}) describes the oscillator 
localization in the sense
discussed above with the average coordinate and momentum subject to random
fluctuation. Equations of such type can be of interest in the theory of
quantum measurement as an alternative to the stochastic Schr\"{o}dinger
equations \cite{Weiss}. In particular, compared to the latter equations,
equation (\ref{gaueq}) does not conserve purity of the state. Such behavior 
is more physically plausible for a system coupled to a bath.

\section{Concluding remarks} \label{conc}

Equilibrium density matrix is often considered as an irreducible concept,
 viz., it is assumed that this is a true state of the bath in each 
individual experiment. 
We believe that such a view is an oversimplification, and it is more
realistic to assume that true state of the bath is a density matrix 
or even a pure state which includes some random component. Ensemble
of such states for different realizations of the random component gives
an equilibrium density matrix. Such a view is partially supported by
classical statistical mechanics. Indeed, most physicists agree that the 
"true state" of the classical bath in thermal equilibrium is a point in 
phase space. The coarse-grained macroscopic observables for such a point 
(e.g., number of particles in a volume element) are essentially the same 
as calculated for one of the Gibbs ensembles (of such points) for large 
enough course-graining \cite{Lebowitz}.  

It is true that the ensemble decomposition of the equilibrium density 
matrix in terms 
of other density matrices (or pure states) is not unique, and that the 
averages calculated for any decomposition are identical to equilibrium 
averages. We believe, however, that it can be of importance, that, when
quantities like variances and $Tr\rho^2$ are calculated for individual 
members of the ensemble, we can get drastically different behavior for
the system for different decompositions, as was shown in this paper.  

\section{Acknowledgments}
The author would like to thank Dr. Gonzalo Ord\'{o}\~{n}ez for
fruitful discussions.
Part of this research was supported by grants from the Robert A. Welch 
Foundation (W-1442) and the Petroleum Research Fund, 
administered by the American Chemical Society.

\appendix*  \label{free}
\section{}
In this Appendix we give explicit expressions for $Tr\rho^2(t)$ for two
types of initial pure states of the oscillator for the RWA Hamiltonian
 if the bath modes are initially in coherent states.

Let the oscillator be initially in a squeezed state obtained by acting on 
the oscillator ground state with the squeezing and displacement operators
\be
|\psi\ket=D(\alpha)S(\varepsilon)|0\ket.
\ee
Here the squeezing operator $S(\varepsilon)$ is given by
\be
S(\varepsilon)
=e^{\frac{\varepsilon^*}{2}a-\frac{\varepsilon}{2}a^{\dag}},
\ee
and $\varepsilon=re^{2i\phi}$ is a complex squeezing parameter.
Using expression (\ref{encoh}) for $Tr\rho^2$ we obtain
\be
Tr\rho^2(t)=\frac{1}{\sqrt{1+4|A(t)|^2(1-|A(t)|^2)\sinh^2r}}.
\ee
The maximum entanglement [or minimum $Tr\rho^2(t)$]  is reached 
when $|A(t)|^2=1/2$. Moreover,
$Tr\rho^2(t)$ is a symmetric function of $|A(t)|^2$ with respect to 
this point.
When $|A(t)|^2=1/2$ we have
\be
Tr\rho^2=\frac{1}{\sqrt{1+\sinh^2r}}.
\ee
We have larger entanglement for larger squeezing with $Tr\rho^2$ going
to zero when $r$ goes to infinity.

We now consider an initial state of the oscillator given by a 
superposition of two coherent states
\be
|\psi\ket=\frac{1}{\sqrt{N}}\bigl(|\alpha\ket+|\beta\ket\bigr),
\ee
where $N$ is the normalization constant,
\be
N=2+\bra\alpha|\beta\ket+\bra\beta|\alpha\ket.\label{no}
\ee
In this case we obtain for $Tr\rho^2(t)$
\bea
Tr\rho^2(t)&=&1+\frac{2}{N^2}\bigl(e^{-(1-|A(t)|^2)|\alpha-\beta|^2}
+e^{-|A(t)|^2|\alpha-\beta|^2} \non \\
& &-e^{-|\alpha-\beta|^2}-1\bigr).
\eea
One can easily verify that $Tr\rho^2(t)$ takes its minimum value
again at $|A(t)|^2=1/2$, and $Tr\rho^2(t)$ is again a symmetric function 
of $|A(t)|^2$ with respect to that point, which may be a general 
property of this model.
Let us consider $Tr\rho^2(t)$ at  $|A(t)|^2=1/2$. We have
\be
Tr\rho^2=1-\frac{2}{N^2}\left(1-e^{-\frac{|\alpha-\beta|^2}{2}}\right).
\ee
Using (\ref{no}), writing $\bra\alpha|\beta\ket$ as $Re^{i\varphi}$ and
remembering that 
\be
e^{-\frac{|\alpha-\beta|^2}{2}}=\left|\bra\alpha|\beta\ket\right|,
\ee
we obtain
\be
Tr\rho^2=1-\frac{(1-R)^2}{2(1+R\cos\varphi)^2}.
\ee
We can see from the last expression that the minimum $Tr\rho^2$ 
decreases with the decreasing overlap between the two coherent states. 
Note, however, that in this case $Tr\rho^2$ 
can never become less then $1/2$.




\begin{thebibliography}{18}
\expandafter\ifx\csname natexlab\endcsname\relax\def\natexlab#1{#1}\fi
\expandafter\ifx\csname bibnamefont\endcsname\relax
  \def\bibnamefont#1{#1}\fi
\expandafter\ifx\csname bibfnamefont\endcsname\relax
  \def\bibfnamefont#1{#1}\fi
\expandafter\ifx\csname citenamefont\endcsname\relax
  \def\citenamefont#1{#1}\fi
\expandafter\ifx\csname url\endcsname\relax
  \def\url#1{\texttt{#1}}\fi
\expandafter\ifx\csname urlprefix\endcsname\relax\def\urlprefix{URL }\fi
\providecommand{\bibinfo}[2]{#2}
\providecommand{\eprint}[2][]{\url{#2}}

\bibitem[{\citenamefont{Caldeira and Leggett}(1983)}]{Caldeira}
\bibinfo{author}{\bibfnamefont{A.~O.} \bibnamefont{Caldeira}} 
\bibnamefont{and}
  \bibinfo{author}{\bibfnamefont{A.~J.} \bibnamefont{Leggett}},
  \bibinfo{journal}{Physica A} \textbf{\bibinfo{volume}{121}},
  \bibinfo{pages}{587} (\bibinfo{year}{1983}).

\bibitem[{\citenamefont{Haake and Reibold}(1985)}]{Haake}
\bibinfo{author}{\bibfnamefont{F.}~\bibnamefont{Haake}} \bibnamefont{and}
  \bibinfo{author}{\bibfnamefont{R.} \bibnamefont{Reibold}},
  \bibinfo{journal}{Phys. Rev. A} \textbf{\bibinfo{volume}{32}},
  \bibinfo{pages}{2462} (\bibinfo{year}{1985}).

\bibitem[{\citenamefont{Grabert et~al.}(1988)
\citenamefont{Grabert, Schramm, and Ingold}}]{Grabert2}
\bibinfo{author}{\bibfnamefont{H.} \bibnamefont{Grabert}},
  \bibinfo{author}{\bibfnamefont{P.} \bibnamefont{Schramm}},
\bibnamefont{and}
  \bibinfo{author}{\bibfnamefont{G.~L.}~\bibnamefont{Ingold}},
  \bibinfo{journal}{Phys. Rep} \textbf{\bibinfo{volume}{168}},
  \bibinfo{pages}{115} (\bibinfo{year}{1988}).

\bibitem[{\citenamefont{Walls and Milburn}(1994)}]{Walls}
\bibinfo{author}{\bibfnamefont{D.~F.}~ \bibnamefont{Walls}} 
\bibnamefont{and}
  \bibinfo{author}{\bibfnamefont{G.~J.}~ \bibnamefont{Milburn}},
  \emph{\bibinfo{title}{Quantum Optics}} 
(\bibinfo{publisher}{Springer-Verlag}, \bibinfo{address}{Berlin},
  \bibinfo{year}{1994}).

\bibitem[{\citenamefont{Carmichael}(1993)}]{Carmichael}
\bibinfo{author}{\bibfnamefont{H.} \bibnamefont{Carmichael}},
  \emph{\bibinfo{title}{An Open System Approach to Quantum Optics}} 
(\bibinfo{publisher}{Springer-Verlag}, \bibinfo{address}{Berlin},
  \bibinfo{year}{1993}).

\bibitem[{\citenamefont{Unruh and Zurek}(1989)}]{Unruh}
\bibinfo{author}{\bibfnamefont{W.~G.} \bibnamefont{Unruh}} 
\bibnamefont{and}
  \bibinfo{author}{\bibfnamefont{W.~H.} \bibnamefont{Zurek}},
  \bibinfo{journal}{Phys. Rev. D} \textbf{\bibinfo{volume}{40}},
  \bibinfo{pages}{1071} (\bibinfo{year}{1989}).

\bibitem[{\citenamefont{Hu et~al.}(1992)
\citenamefont{Hu, Paz, and Zhang}}]{Hu}
\bibinfo{author}{\bibfnamefont{B.~L.} \bibnamefont{Hu}},
  \bibinfo{author}{\bibfnamefont{J.~P.} \bibnamefont{Paz}}, 
\bibnamefont{and}
  \bibinfo{author}{\bibfnamefont{Y.}~\bibnamefont{Zhang}},
  \bibinfo{journal}{Phys. Rev. D} \textbf{\bibinfo{volume}{45}},
  \bibinfo{pages}{2843} (\bibinfo{year}{1992}).


\bibitem[{\citenamefont{Guilini et~al.}(1996)\citenamefont{Guilini, Joos,
  Kiefer, Kupsch, Stomatescu, and Zeh}}]{Guilini}
\bibinfo{author}{\bibfnamefont{D.}~\bibnamefont{Guilini}},
  \bibinfo{author}{\bibfnamefont{E.}~\bibnamefont{Joos}},
  \bibinfo{author}{\bibfnamefont{C.}~\bibnamefont{Kiefer}},
  \bibinfo{author}{\bibfnamefont{J.}~\bibnamefont{Kupsch}},
  \bibinfo{author}{\bibfnamefont{I.~O.} \bibnamefont{Stomatescu}},
  \bibnamefont{and} \bibinfo{author}{\bibfnamefont{H.~D.} 
\bibnamefont{Zeh}},
  \emph{\bibinfo{title}{Decoherence and the Appearance of 
a Classical World in Quantum Theory}} (\bibinfo{publisher}{Springer}, 
\bibinfo{address}{Berlin},
  \bibinfo{year}{1996}).

\bibitem[{\citenamefont{Hakim and Ambegaokar}(1985)}]{Hakim}
\bibinfo{author}{\bibfnamefont{V.}~\bibnamefont{Hakim}} \bibnamefont{and}
  \bibinfo{author}{\bibfnamefont{V.} \bibnamefont{Ambegaokar}},
  \bibinfo{journal}{Phys. Rev. A} \textbf{\bibinfo{volume}{32}},
  \bibinfo{pages}{423} (\bibinfo{year}{1985}).

\bibitem[{\citenamefont{Karrlein and Grabert}(1997)}]{Grabert}
\bibinfo{author}{\bibfnamefont{R.}~\bibnamefont{Karrlein}} 
\bibnamefont{and}
  \bibinfo{author}{\bibfnamefont{H.} \bibnamefont{Grabert}},
  \bibinfo{journal}{Phys. Rev. E} \textbf{\bibinfo{volume}{55}},
  \bibinfo{pages}{153} (\bibinfo{year}{1997}).

\bibitem[{\citenamefont{Romero and Paz}(1997)}]{Romero}
\bibinfo{author}{\bibfnamefont{L.~D.}~\bibnamefont{Romero}}
\bibnamefont{and}
  \bibinfo{author}{\bibfnamefont{J.~P.} \bibnamefont{Paz}},
  \bibinfo{journal}{Phys. Rev. A} \textbf{\bibinfo{volume}{55}},
  \bibinfo{pages}{4070} (\bibinfo{year}{1997}).

\bibitem[{\citenamefont{da Costa et~al.}(2000)
\citenamefont{da Costa,Caldeira,Dutra, and Westfahl}}]{daCosta}
\bibinfo{author}{\bibfnamefont{M.~R.} \bibnamefont{daCosta}},
  \bibinfo{author}{\bibfnamefont{A.~O.} \bibnamefont{Caldeira}},
\bibinfo{author}{\bibfnamefont{S.~M.} \bibnamefont{Dutra}},
\bibnamefont{and}
  \bibinfo{author}{\bibfnamefont{H.}~\bibnamefont{Westfahl}},
  \bibinfo{journal}{Phys. Rev. A} \textbf{\bibinfo{volume}{61}},
  \bibinfo{pages}{022107} (\bibinfo{year}{2000}).

\bibitem[{\citenamefont{Petrosky and Barsegov}(2002)}]{Petrosky}
\bibinfo{author}{\bibfnamefont{T.}~\bibnamefont{Petrosky}} 
\bibnamefont{and}
  \bibinfo{author}{\bibfnamefont{V.} \bibnamefont{Barsegov}},
  \bibinfo{journal}{Phys. Rev. E} \textbf{\bibinfo{volume}{65}},
  \bibinfo{pages}{046102} (\bibinfo{year}{2002}).

\bibitem[{\citenamefont{Hu and Matacz}(1994)}]{Hu2}
\bibinfo{author}{\bibfnamefont{B.~L.}~\bibnamefont{Hu}}
\bibnamefont{and}
  \bibinfo{author}{\bibfnamefont{A.} \bibnamefont{Matacz}},
  \bibinfo{journal}{Phys. Rev. D} \textbf{\bibinfo{volume}{49}},
  \bibinfo{pages}{6612} (\bibinfo{year}{1994}).

\bibitem[{\citenamefont{Bogoliubov and Bogoliubov}(1994)}]{Bogoliubov}
\bibinfo{author}{\bibfnamefont{N.~N.} \bibnamefont{Bogoliubov}}
\bibnamefont{and}
  \bibinfo{author}{\bibfnamefont{N.~N.} \bibnamefont{Bogoliubov, Jr.}},
  \emph{\bibinfo{title}{An Introduction to Quantum Statistical Mechanics}}
(\bibinfo{publisher}{Gordon and Breach Science Publishers},
\bibinfo{address}{Philadelphia},
  \bibinfo{year}{1994}).

\bibitem[{\citenamefont{Ford et~al.}(1965)
\citenamefont{Ford,Kac, and Mazur}}]{Ford}
\bibinfo{author}{\bibfnamefont{G.~W.} \bibnamefont{Ford}},
  \bibinfo{author}{\bibfnamefont{M.} \bibnamefont{Kac}},
\bibnamefont{and}
  \bibinfo{author}{\bibfnamefont{P.}~\bibnamefont{Mazur}},
  \bibinfo{journal}{J. Math. Phys.} \textbf{\bibinfo{volume}{6}},
  \bibinfo{pages}{504} (\bibinfo{year}{1965}).

\bibitem[{\citenamefont{Lindenberg and West}(1984)}]{Katja}
\bibinfo{author}{\bibfnamefont{K.}~\bibnamefont{Lindenberg}} 
\bibnamefont{and}
  \bibinfo{author}{\bibfnamefont{B.~J.} \bibnamefont{West}},
  \bibinfo{journal}{Phys. Rev. A} \textbf{\bibinfo{volume}{30}},
  \bibinfo{pages}{568} (\bibinfo{year}{1984}).


\bibitem[{\citenamefont{Ford et~al.}(1988)
\citenamefont{Ford,Lewis, and O'Connell}}]{Ford2}
\bibinfo{author}{\bibfnamefont{G.~W.} \bibnamefont{Ford}},
  \bibinfo{author}{\bibfnamefont{J.~T.} \bibnamefont{Lewis}},
\bibnamefont{and}
  \bibinfo{author}{\bibfnamefont{R.~F.}~\bibnamefont{O'Connell}},
  \bibinfo{journal}{Phys. Rev. A} \textbf{\bibinfo{volume}{37}},
  \bibinfo{pages}{4419} (\bibinfo{year}{1988}).


\bibitem[{\citenamefont{Cahill and Glauber}(1969)}]{Glauber10}
\bibinfo{author}{\bibfnamefont{K.~E.}~\bibnamefont{Cahill}} 
\bibnamefont{and}
  \bibinfo{author}{\bibfnamefont{R.~J.} \bibnamefont{Glauber}},
  \bibinfo{journal}{Phys. Rev. } \textbf{\bibinfo{volume}{177}},
  \bibinfo{pages}{1857} (\bibinfo{year}{1969}).

\bibitem[{\citenamefont{Gradshteyn and Ryzhik}(1980)}]{Gradshtein}
\bibinfo{author}{\bibfnamefont{I.~S.} \bibnamefont{Gradshteyn}} 
\bibnamefont{and}
  \bibinfo{author}{\bibfnamefont{I.~M.} \bibnamefont{Ryzhik}},
  \emph{\bibinfo{title}{Tables of Integrals, Series, and Products}} 
(\bibinfo{publisher}{Academic Press}, \bibinfo{address}{New York},
  \bibinfo{year}{1980}).

\bibitem[{\citenamefont{Glauber}(1966)}]{Glauber3}
\bibinfo{author}{\bibfnamefont{R.~J.} \bibnamefont{Glauber}},
  \bibinfo{journal}{Phys. Lett.} \textbf{\bibinfo{volume}{21}},
  \bibinfo{pages}{650} (\bibinfo{year}{1966}).

\bibitem[{\citenamefont{Weiss}(1999)}]{Weiss}
\bibinfo{author}{\bibfnamefont{U.}~\bibnamefont{Weiss}},
  \emph{\bibinfo{title}{Quantum Dissipative Systems}}
  (\bibinfo{publisher}{World Scientific}, \bibinfo{address}{Singapore},
  \bibinfo{year}{1999}), \bibinfo{edition}{2nd} ed.

\bibitem[{\citenamefont{Lebowitz}(1993)}]{Lebowitz}
\bibinfo{author}{\bibfnamefont{J.~L.} \bibnamefont{Lebowitz}},
  \bibinfo{journal}{Physica A} \textbf{\bibinfo{volume}{194}},
  \bibinfo{pages}{1} (\bibinfo{year}{1993}).



\end{thebibliography}
\end{document}